\documentclass[letterpaper, 10 pt, conference]{ieeeconf}  

\IEEEoverridecommandlockouts                              

\usepackage{graphicx} 
\graphicspath{ {Images/} }
\usepackage{amsmath} 
\usepackage{amssymb}  
\usepackage{flushend}
\usepackage{url}
\usepackage[dvipsnames]{xcolor}
\title{\LARGE \bf
Implementation of a Natural User Interface to Command a Drone
}

\author{Yam-Viramontes, Brandon. A.$^{1}$ \and Mercado-Ravell, Diego.$^{2}$
\thanks{*This work was not supported by any organization}
\thanks{$^{1}$Yam-Viramontes, B. A is with the Instituto Tecnologico Superior de Jerez ITSJ, Zacatecas, Mexico.
        {\tt\small email: abelnight5057@gmail.com}}%
\thanks{$^{2}$D. Mercado-Ravell is with Cátedras CONACYT at the Research Center in Mathematics CIMAT-Zac, Unit Zacatecas, Mexico, (corresponding author) phone: +52-449-428-4800;
        {\tt\small email: diego.mercado@cimat.mx}}%
}

\begin{document}

\maketitle
\thispagestyle{empty}
\pagestyle{empty}

\begin{abstract}
 In this work, we propose the use of a Natural User Interface (NUI) through body gestures using the open source library OpenPose, looking for a more dynamic and intuitive way to control a drone. For the implementation, we use the Robotic Operative System (ROS) to control and manage the different components of the project. Wrapped inside ROS, OpenPose (OP) processes the video obtained in real-time by a commercial drone, allowing to obtain the user's pose. Finally, the keypoints from OpenPose are obtained and translated, using geometric constraints, to specify high-level commands to the drone.  Real-time experiments validate the full strategy.

\end{abstract}

\section{INTRODUCTION}

In recent years the constant innovation in the field of drones has allowed them to be more affordable, increase their autonomy and capabilities, augmenting the number of tasks favoring their application in areas such as shipping and delivery \cite{delivery}, crowd monitoring \cite{crowd}, precision agriculture \cite{agriculture}, geography mapping \cite{mapping}, aerial photography \cite{aereo}, entertainment \cite{entertainment}, etc.

Nowadays it's common to see people without  knowledge on the subject who have their own drone, either to perform a certain task or simply for hobby. For this reason, the economic sector has focused on developing more friendly systems. Thanks to that, most of the quadrotors in the market are semi-autonomous, and their control is accomplished through an application or radio-control. This can change with the help of the Natural User Interfaces (NUI's), which are defined as a system for human-computer interaction where the user operates through intuitive actions related to natural, everyday human behavior. For example, through body gestures and voice commands.

According to the state of art, most of the results seem to indicate that the implementation of a NUI facilitates the Human-Drone Interaction (HDI) in simple tasks in order to make them more intuitive and user-friendly.
Probably for this reason, in recent years its application looks more palpable in several areas. For example, within the health sector, we find the case of a virtual training system that encourages the user to perform physical activities in a motivating and entertaining way \cite{jofre}, we also find its application in the development of a telemedicine platform that facilitates the rehabilitation of people with hip prostheses \cite{tiago}. In the technological sector, applications have been made in the areas of interaction with robots and virtual reality environments. For example in \cite{c1}, an exploratory study was carried out to investigate gestures with speech interfaces for interaction with robots in a simulation augmented reality, such is the case of the proposal of different gesture recognition techniques in a Leap Motion Controller Augmented reality framework. \cite{c2}. 

    \begin{figure}[t!]
      \centering
      \includegraphics[width=0.48\textwidth]{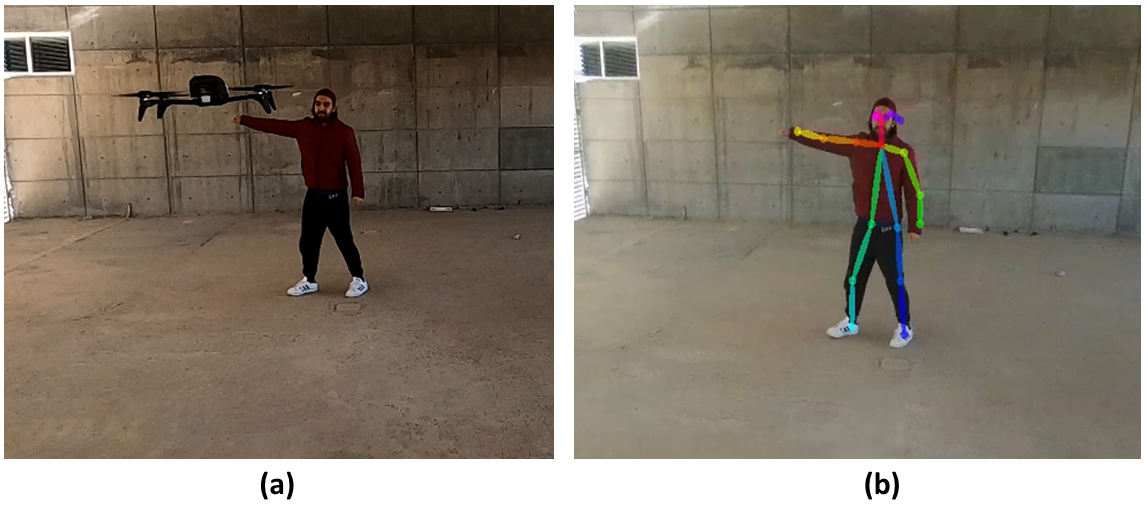}

      \caption{Natural User Interface (NUI) to control a drone using body gestures. The use of NUIs allows improving the human-drone interaction experience, making the task more intuitive and entertaining. (a) Quadcopter following high-level commands from body gestures. (b) User's body pose obtained by the OpenPose algorithm.}
      \label{fig_prueba}
   \end{figure}
   
More specific to the field of drones, other related works have been carried out, for example in \cite{c4} the response of four different NUI methods implemented in a drone was tested. Those were: tracking a human user, performing command tasks in base of visual markers and hand gestures, and finally speech commands. In \cite{c3}, a method to control a drone directed by the facial pose was developed. Also in \cite{c5}, visual detection and tracking with drones were implemented using ROS, in order to follow and interact with the human user, developing more robust control by adding Kalman's filters. In \cite{c6}, a drone control was proposed using a kinect sensor, obtaining points of interest and using control gestures. In all of these cases, the user's response, in general, was positive. However, most of them rely on special instruments or external devices to identify body gestures. Closer to our work, in \cite{c7} a NUI was created to recognize specific user positions. They obtained image regions and joint positions of human bodies in images through OP and then the feature vectors of a human body were generated and classified by a Support Vector Machine (SVM). This classifier allows recognizing only four different postures of the user. They used ROS, but only to analyze the performance of their proposal through ``bagfiles" in order to record image data and to corroborate your results. This method has good an average accuracy and consumes limited computing resources.

Our current proposal is a NUI that allows sending high-level commands to the drone in real-time, through hand and arm gestures obtained from OP, being able to recognize ten different postures while consuming  fewer resources, furthermore, it is not necessary to buy some extra component for its implementation in drone, as depicted in Fig.   \ref{fig_prueba} . All the algorithms run in real-time in a ground station computer with the help of ROS. Real-time experiments validate our proposed interface, allowing the user to control the drone in an intuitive and more entertaining manner.

The outline of this paper is the following: The NUI concept is further developed in Section \ref{NUI}, explaining how our proposal is composed and how it works. In Section \ref{setup}, the configuration of our experimental platform is explained. The experimental results of our proposal are described in Section \ref{results}. Finally in Section \ref{conclusions} we discuss our conclusion and future work.
   
\section{NATURAL USER INTERFACE}
\label{NUI}
Natural user interfaces (NUIs) are the next step in the evolution of user interfaces. Compared to Graphical User Interfaces (GUI), which have been characterized by providing a more intuitive interface that helps to reduce the barrier between the user and the machine, the NUIs opt for a control centered on the characteristic gestures of the human being, being able to summarize these characteristics in the definition given in \cite{nuidefinition} that tells us about them as: ``a kind of interface that enables users to interact with computing devices in the same way they interact with the physical world, through using voice, hands, and body movements".  Today we can find many of its implicit applications in our daily lives, sometimes without realizing that we are interacting with one of them. Some examples of these cases are found in the voice assistants currently on the market,  as Google assistant, Alexa, and Siri that respond to a voice-controlled NUI \cite{nuiassistants}. Another common implementation is in video game console components \cite{nuigames} as the Wiimote, Kinect, sony move controller, etc., allowing to make the game more immersive.

Although in recent years there has been a great development in this area it is still necessary to continue improving these interfaces, creating more robust systems, as well as defining more intuitive poses for the user. At present, works focusing on this topic have been developed as in \cite{dataset} which proposes a dataset which is composed of 13 gestures suitable for basic unmanned aerial vehicles navigation and command from general aircraft and helicopter signals. The dataset proposed in \cite{dataset2} includes the following actions, side-view actions and front-view actions, also composed of 13 concrete actions. Another very interesting work presented in  \cite{dronme} is the application of an experiment called the Wizard of Oz (WoZ) which is used to determine what are the postures and body gestures that users rate as more comfortable to command a drone through a NUI. This experiment is based on subjects that interact with a computer system that they believe to be autonomous, but which is actually being fully or partially operated by an unseen human being \cite{woz}, similar to the famous fairy tail.

We propose to implement a NUI to control a drone, using only information embedded on the vehicle and a ground station, so that it helps inexperienced users to easily command it using body gestures as explained in the following.

    \begin{figure}[t!]
    \label{fig_keypoints}
      \centering
      \includegraphics[width=0.3\textwidth]{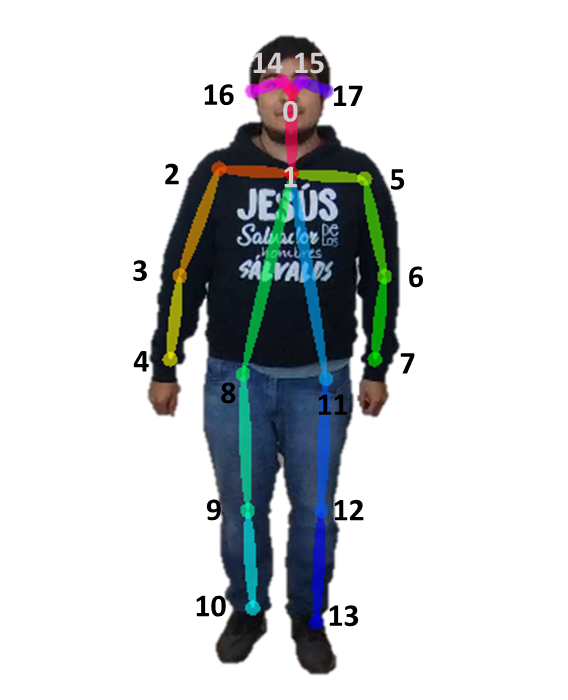}
      \caption{Location of the 18 keypoints defining the user's pose, provided by OpenPose.}
   \end{figure}

\subsection{Implementation}
For our proposal, we decided not to depend on any additional external device or marker, such as a Kinect or some depth camera. Instead, we decided to use a pose detector, so that the only thing necessary is images of the user in real-time  provided by the cameras usually included with drones. Within the state of the art, we find very interesting proposals, such as OpenPose \cite{openpose} which is a real-time multi-person keypoint detection library, Pose Proposal Networks \cite{ppn} which is a method to detect an unknown number of articulated 2D poses in real-time, DensePose \cite{densepose} which maps all human pixels of an RGB image to the 3D surface of the human body for pose estimation and wrnchAI \cite{AI} that is a frictionless motion capture and activity recognition system. However, among these options, we discard wrnchAI despite it consumes less resources and makes faster detections since a license is required for its use. On the other hand, \cite{ppn} focuses on the estimation of the 2D pose of several people from a 2D still image. Therefore, not knowing its performance when processing a video, we disregarded it. Moving forward, Densepose it's more focused on establishing dense correspondences from a 2D image to a 3D, surface-based representation of the human body\cite{densepose}. Even though DensePose seems to offer a suitable alternative for our approach, we opted to use OpenPose \cite{openpose} for its simplicity in the pose's representation, allowing us to identify body gestures using only geometric  constraints. 

Thus, in the development of our NUI, we decided to process images obtained in real-time with OP, which is an open-source real-time system for multi-person 2d pose detection which give us the position of several members of the user's body, each represented in two $x, y$ coordinates, and a third value indicating the confidence score of the detection in a range between $[0,1]$. Each of these members is numbered from 0 to 17, as depicted in Fig. \ref{fig_keypoints}.

For our proposal, for simplicity as a first stage, we consider to only focus on the user's upper extremities, Therefore, we consider only 6 points of interest, which are mentioned in Table \ref{table_specific_points}. We have also forced the detection algorithm to detect only one person at a time, in order to improve computational efficiency.

\begin{table}[b]
\caption{Specific points used in our implementation}
\label{table_specific_points}
\begin{center}
\begin{tabular}{|c||c|}
\hline
nose (0)& right wrist (4)\\
\hline
neck (1) & left shoulder (5)\\
\hline
right shoulder (2)& left wrist (6)\\
\hline
\end{tabular}
\end{center}
\end{table}

\begin{figure}[t]
      \centering
      \includegraphics[width=0.3\textwidth]{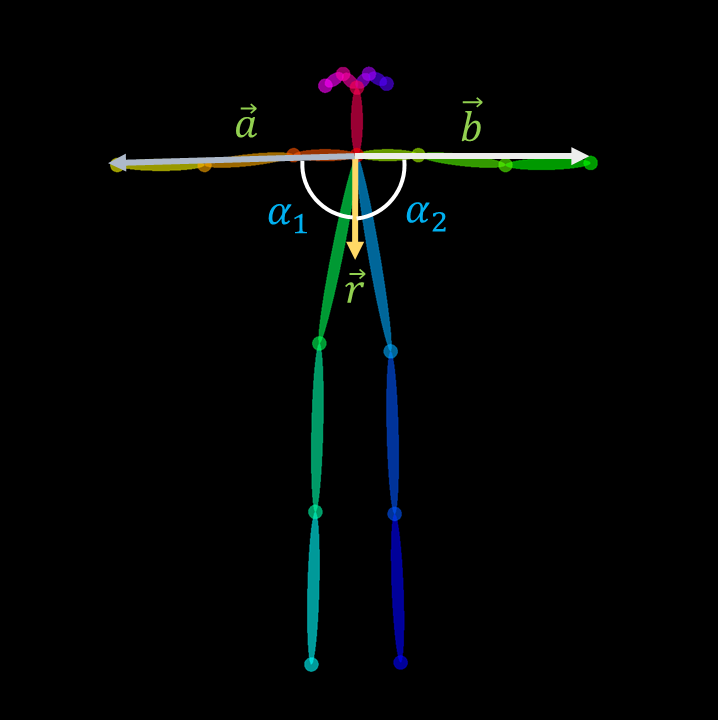}
            \caption{Representation of considered vectors and nomenclature of the angles formed between them.}
      \label{fig_angles}
   \end{figure}

These points are processed using a python script, aiming to translate specific angles and normalized distances into high order commands to the drone, using only simple geometry  constraints.
To achieve this, we read the different points in the array provided by OP, from the input image. As mentioned earlier, the python script is responsible for calculating two elements, angles, and normalized distances. For the first of them, we decided to consider some specific points as vectors. By default, the image considers the point of origin located in the upper left corner (according to the rule of the right hand) therefore, seeking to make it more intuitive, we define a body reference frame centered point $1$, corresponding to the user chest center. Now from this point, we will consider 3 vectors, one for reference ($\vec{r}$) and two that will define the angle formed ($\vec{a}$ and $\vec{b}$), as depicted in Fig. \ref{fig_angles}.
  
Therefore, to obtain the angle $\alpha_{1}$, we apply: 
\begin{equation}
    \cos \alpha_{1} = \frac{\overrightarrow{a}.\overrightarrow{r}}{|\overrightarrow{a}||\overrightarrow{r}|}\
\end{equation}
similarly, the same applies to obtain $\alpha_{2}$ with its corresponding vector. Commands respond to the combination of angles within certain intervals which are described later.

we also use normalized distances between points to define body gestures. The use of normalized distances helps to make the algorithm more robust against scale variations due to perspective or different user sizes.

\begin{figure}[t]
      \centering
      \includegraphics[width=0.3\textwidth]{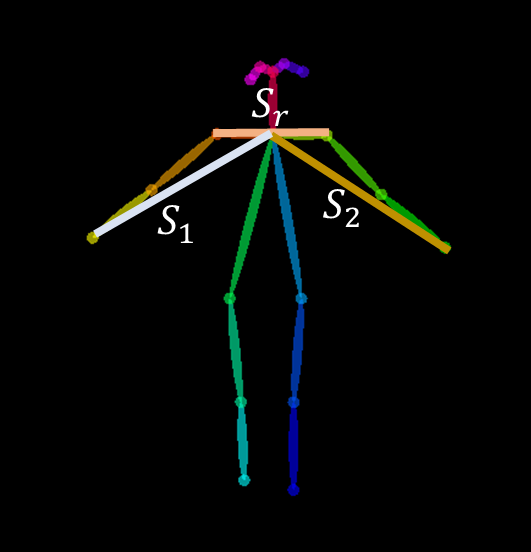}
      \caption{Representation of defined normalized distances and their corresponding nomenclature.}
      \label{fig_proportion}
   \end{figure}

As with the angles, for the normalized distances, we also take one reference distance value which is given by the distance between two points ($S_{r}$) corresponding to both shoulders. Two more distances are also used, called $S_{1}$ and $S_{2}$ respectively, depicted in Fig. \ref{fig_proportion}.

In this case, gestures are defined using only geometric  constraints between the angles and normalized distances defined before. Table \ref{table_commands_drone} summarizes the different body gesture commands along with their respective definition in terms of the 5 different parameters, angles, and distances, where NA means that in that case that condition is not considered.  

A  total  of  10  body gestures corresponding  to different  high-level  commands  are  considered.  ``Left”  and "right” commands are defined by the corresponding extended arm elevation, up to shoulder height. For the ``up” command,both  arms  must  be  above  the  shoulders,  analogously,  both arms must be below the shoulder at about 45 degrees for the ``down” command. Also, for yaw rotation, one arm must be above the shoulder, which will define the direction of movement, either ``turn clockwise (CW)” or ``turn counterclockwise (CCW)”, while the other arm must maintain a certain elevation below  the  shoulder.  For  the  "forward”  command,  the  right arm  should  be  near  the  face  while  the  left  is  at  rest.  If  the left arm is used instead, the "backwards” command is sent. Additionally,  To  take  a  picture  of  the  user,  both  arms  must be  close  to  the  face.  Finally,  the  "wait”  command  is  sent when both arms are at rest, commanding the drone to hover in place. These body gestures and their respective commands are presented in Fig. 7.

\begin{table}[b]
Body gesture commands' definition
\label{table_commands_drone}
\begin{center}
\begin{tabular}{|c||c||c||c|}
\hline
command & $\alpha_{1}$ & $\alpha_{1}$ & distance based scales (s)\\
\hline
Snapshot & NA & NA & $S_{1}< S_{r} $ and $S_{2}< S_{r}$\\
\hline
Backward & [0,40) & NA & $S_{2}< S_{r} $ \\
\hline
Forward & NA & [0,40) & $S_{1}< S_{r} $ \\
\hline
Left &  [0,40) & [70,100) & $S_{r}< S_{1}  $ \\
\hline
Right & [70,100) & [0,40) & $S_{r}< S_{2}$  \\
\hline
Up & [80,180) & [80,180] & $S_{r} < S_{1} $ and $S_{r} < S_{2}$  \\
\hline
Down & [40,80) & [40,80) &  $S_{r} < S_{1} $ and $S_{r} < S_{2}$ \\
\hline
Turn cw & [40,85) & [85,180) & $S_{r} < S_{1} $ and $S_{r} < S_{2}$ \\
\hline
Turn ccw & [85,180) & [40,85) & $S_{r} < S_{1} $ and $S_{r} < S_{2}$  \\
\hline
Wait & [0,40) & [0,40) & $S_{r} < S_{1} $ and $S_{r} < S_{2}$  \\
\hline
\end{tabular}
\end{center}
\end{table}

Delving into the structure of our algorithm whit the help of the ROS \cite{ros}, which is a middleware that facilitates the intercommunication of several processes through an organized graph structure. Three main nodes are employed. The first one is the drone's driver which recovers the video from the frontal camera mounted in the drone, along with the rest of the sensor's information. This node also allows us to send high order commands to the vehicle. The second key node receives the video stream from the drone and uses it to implement the OP algorithm, outputting the human body's pose from the main user. The third node translates the human body gestures into high-level commands for the drone, as described previously.
See Fig. \ref{fig_ros}.

A graphic representation of the different arm gestures used for the proposed NUI is depicted in Fig. \ref{fig_gestures}.

In order to reduce the effects of noise and to smooth the commands sent to the drone another security measure implemented,  the command is not sent to the drone until it is repeated two consecutive times in our algorithm. Also, if no command is detected after two consecutive iterations, an instruction is sent so that the drone hovers in place.

\section{ Experiment setup}
\label{setup}

For the development of this project, we had used a commercial drone type bebop 2 from parrot, which has a frontal camera of $14$ megapixels with a grand angular lens and a video resolution of  $1920 \times 1080p$ (30 fps), moreover, a digital image stabilization system is provided. It uses a $2700$ mAh Lithium Potassium (LiPo) battery that allows an autonomy of 25 minutes of flight. It is a small drone, with dimensions $8\times33\times9$cm, and very light-weight with only $500$ grams. Additionally, it is provided with a Software Development Kit (SDK).
As a ground station, we use a laptop computer with operating system Ubuntu 18.04.4, Nvidia 920mx graphics card, 8gb of ram, intel core processor i5-7200U. We also use CUDA 10.0, CUDNN 7.2 and caffe which is a deep learning framework.

Regarding the operation of the drone that is used, we consider the well-known dynamic model of a quadcopter \cite{c9}:

\begin{equation}
        \begin{bmatrix} 
        \ddot{x}\\ 
        \ddot{y}\\ 
        \ddot{z} 
    \end{bmatrix} 
    \approx 
    \begin{bmatrix} 
        s\psi s\phi + c\psi s\theta c\phi \\
        -c\psi s\phi + s\psi s\theta c\phi \\
        c\theta c\phi 
    \end{bmatrix}
    -\begin{bmatrix} 
        0\\ 
        0\\ 
        g 
    \end{bmatrix}
    \label{eq_model}
\end{equation},
\begin{equation}
\begin{bmatrix} \ddot{\Phi }\\ \ddot{\Theta }\\ \ddot{\Psi } \end{bmatrix} \approx \begin{bmatrix} \tau_{\Phi}\\ \tau_{\Theta}\\ \tau_{\Psi} \end{bmatrix}
\end{equation}
where $x, y, z$ are the position of the quadcopter with respect to an inertial frame, $T$ $\in$ $\mathbb{R}$ defines the total thrust produced by the motors. Meanwhile, $m$ and $g$ represent the mass and gravity constant, respectively. $[\phi, \theta, \psi]$ stand for the Euler angles roll, pitch and yaw, and $[\tau_{\phi}, \tau_{\theta}, \tau_{\psi},]$ describe the control torques produced by the differential velocities of the rotors. The short notation $[s\alpha = sin(\alpha );\ c\alpha = cos(\alpha )]$ is used.

\begin{figure}[t]
      \centering
      \includegraphics[width=0.5\textwidth]{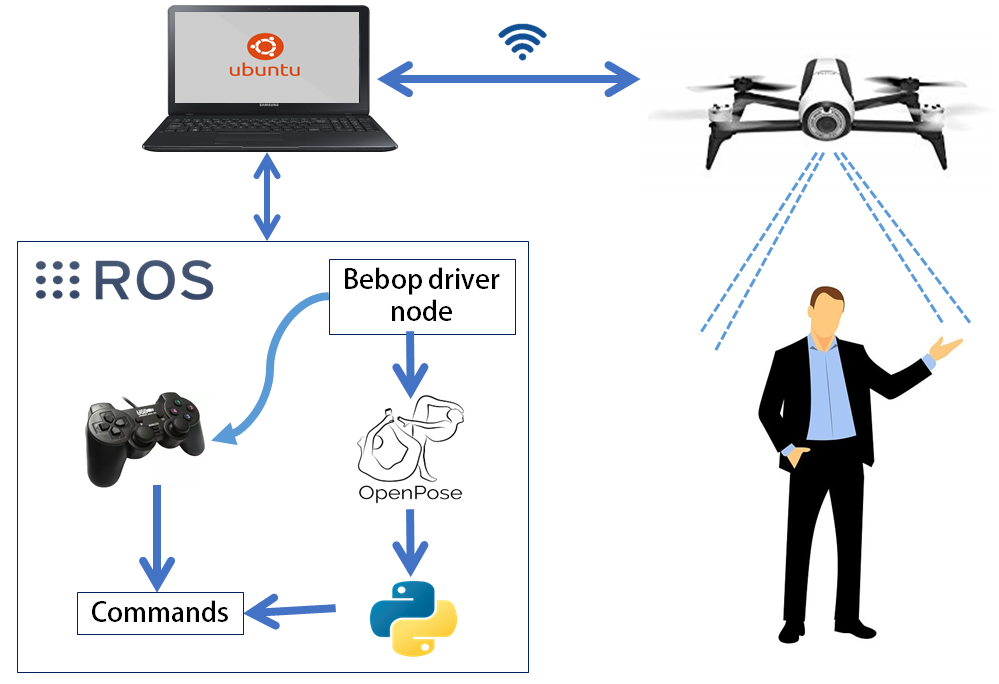}
     
      \caption{Overall system description. The drone communicates wirelessly with a ground station, composed of a computer running ROS. Three main ROS nodes are implemented, one for the drone's driver, one for the OP algorithm and the last one for the NUI's implementation, converting the body poses into high-level commands to the drone. Also, a joystick is available to recover the drone in case of an emergency. }
       \label{fig_ros}
   \end{figure}

\begin{figure}[t!]
      \centering

      \includegraphics[width=0.45\textwidth]{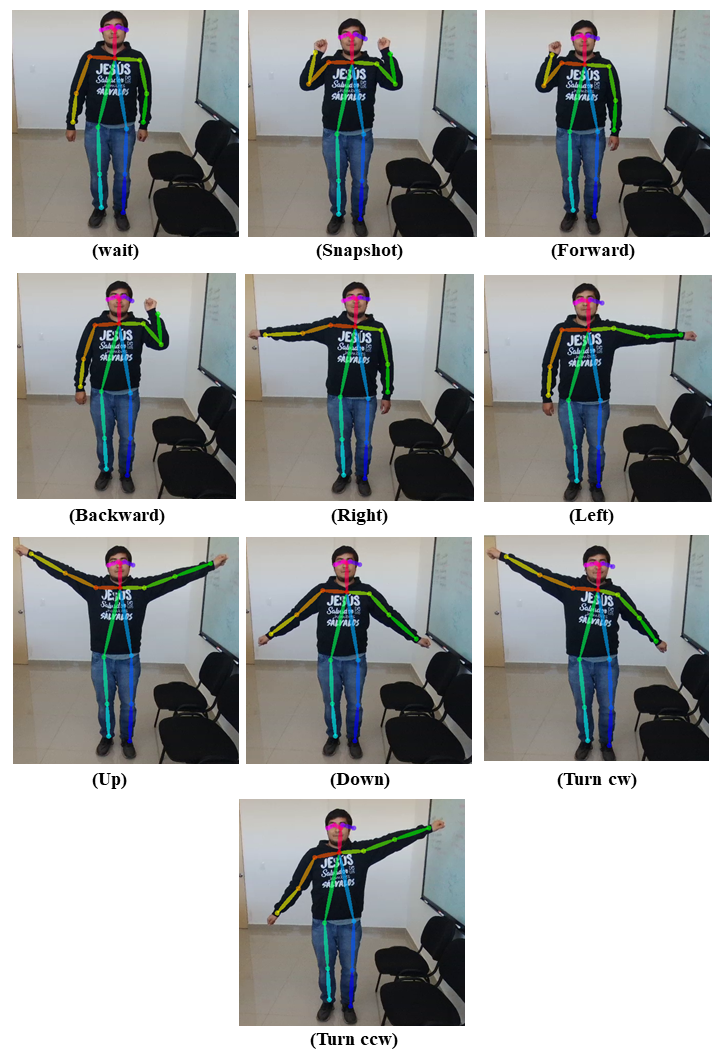}
      
      \caption{Representation of the different body gestures that the user must use as well as their corresponding high-level command sent to the drone.}
      \label{fig_gestures}
   \end{figure}

We use OP wrapped with ROS. ROS has nodes that can send, receive and multiplex messages simultaneously. It is a fundamental component of our NUI, and it's responsible for connecting the three main nodes. An additional node is used s a security measure, to read commands from a joystick, which serves to recover the drone manually if necessary. It also helps to initialize the system and take off the vehicle (see Fig. \ref{fig_ros}).
   
\section{Real-time Experiments}
\label{results}

When working around human users, the user's safety is the main constrain. Henceforth, the tests performed in an indoor environment which allowed us to have controlled conditions, thus counting with greater security. As an extra safety measure, both take-off and landing of the drone were provided manually by a human test supervisor using a joystick.

\begin{figure}[t!]
      \centering

      \includegraphics[width=0.4\textwidth]{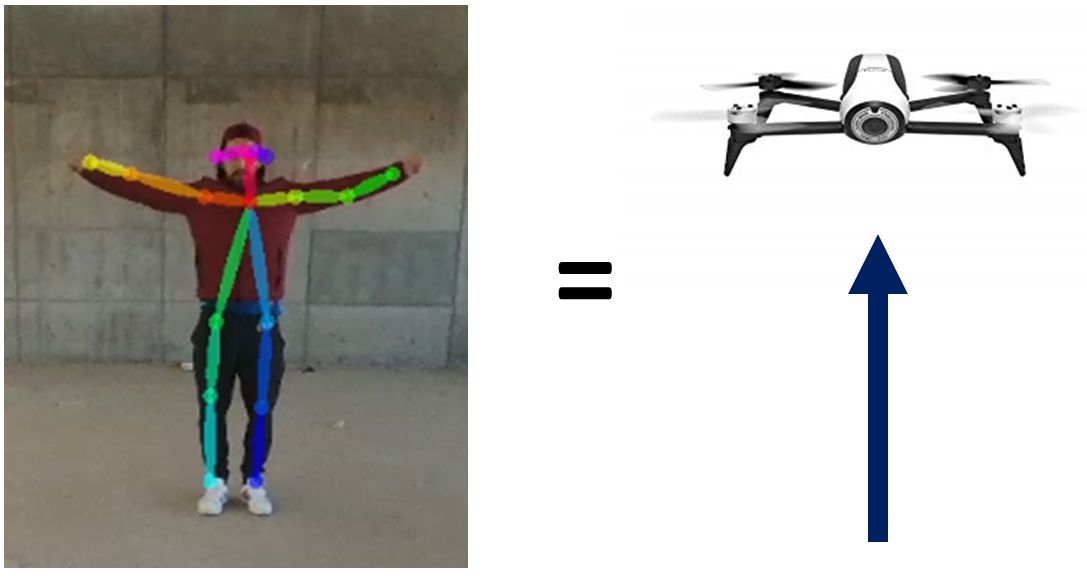}
      
      \caption{Real-time validation of our proposal with a volunteer user in an indoor environment. Here, the human user raises both arms to command the drone to move upwards.}
      \label{fig_commands}
   \end{figure}

For these tests, three volunteers participated, one by one. It is important to notice that two of them did not have previous experience using drones. The tests were divided into two steps. In the first one, the pose required for each command was explained to the user and immediately afterward he was asked to do the test following the instructions of the supervisor. For the second step, being familiar with the commands, the user was given complete freedom to perform the gestures of his/her choice. See Fig. \ref{fig_commands}.

\begin{figure}[t!]
      \centering

      \includegraphics[width=0.45\textwidth]{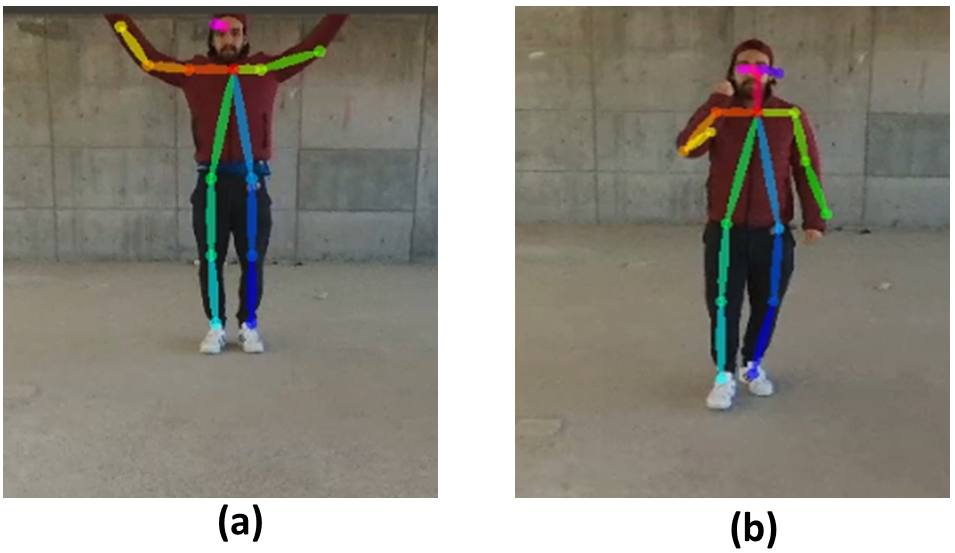}
      
      \caption{Some occasional faults were observed during the development of the tests. In (a) the user leaves the field of view of the drone so the command is not recognized. In (b) the OP algorithm fails to estimate the pose, sometimes giving an incorrect command.}
      \label{fig_errors}
   \end{figure}

At the end of the tests, participants were asked about how was their experience with the drone, with satisfactory feedback, pointing out that it was somewhat entertaining. They also shared the points they didn't like, such as the response time of the drone after giving it a command which currently ranges from $1.5$ to $2.5$ seconds, and having to follow the drone so as not to leave his field of vision. Also, they commented that although some of the gestures were intuitive and comfortable, such as ``Left", ``Right", ``Up" and ``Down", others are not so much, as ``turn CW" and ``turn CCW". Also, participants mentioned that maintaining the arms' position was somehow exhausting in some cases. Moreover, they expressed to feel safer piloting the drone in this way, rather than with a traditional method, and that they perceived it more friendly. Finally, all three volunteers agreed that the overall experience was something that they would like to try more thoroughly.

Although the user experiences were improved using our proposed strategy, some problems were observed occasionally during the tests, compromising to some degree the user interaction, when he/she got to perceive them. Two of these problems are illustrated in Fig. \ref{fig_errors}, and are directly related to user detection. More in particular, the OP algorithm may provide an incorrect body pose estimation, especially if the user goes out of the camera's field of view, or when he/she is too far away from the drone. Both problems can be significantly mitigated by adding an algorithm to track and follow the target user in order to keep a valid distance with the drone (\cite{c5}).

In general, we consider that the overall performance of the proposed system meets the goal of being more intuitive, entertaining and easy to use, in spite of the user's experience using drones.

The video of our experimental validation can be found at: \url{https://youtu.be/0sfcfzFEBdw}

\section{Conclusions and Future Work}
\label{conclusions}

The main advantage of our system is that through simple tools and with hardware with mid-range specifications, we have been able to implement a simple but effective NUI using body gestures, capable of responding to 10 different commands. Furthermore, from the experiments, it is suggested that the implementation of a NUI greatly improves the user experience in handling a drone, even if the user has never used one.

In the future, it is planned to refine the gestures such that they are more intuitive for the user, for example using the ones reported in the literature at \cite{dataset,dataset2,dronme}, and performing tests with more volunteers to determine the degree of comfort during the experience.

Although the fully experience resulted entertaining to the participants, the delays found in the drone's response compromised the interaction. Hence, we aim to improve the system performance by testing lighter algorithms, or improving the current hardware, allowing an even more fluid performance.

 Furthermore, the use of lighter algorithms would allow us to implement our proposal embedded in the vehicle, such that a ground station is not required to run all the processes. Also, it is desired  to implement user tracking, thus preventing the user from abandoning the drone's field of view, and allowing a greater sense of freedom for the user. 

\addtolength{\textheight}{-12cm}   

\end{document}